\documentclass[sigconf]{acmart}
\usepackage{fancyhdr}
\usepackage{xspace}

\AtBeginDocument{%
  }

\setcopyright{acmlicensed}
\copyrightyear{2024}
\acmYear{2024}
\acmDOI{XXXXXXX.XXXXXXX}

\acmConference[ACM CCS HealthSec 2024]{October 14th, 2024}{Salt Lake City, Utah}
\newif\ifdraft
\draftfalse

\newif\ifpreprint
\preprinttrue
\ifdraft
  \newcommand{\todocolor}[1]{\textcolor{red}{#1}}
  \newcommand{\todocolorb}[1]{\textcolor{teal}{#1}}
  \newcommand{\todocolorc}[1]{\textcolor{blue}{#1}}
\else
  \newcommand{\todocolor}[1]{}
  \newcommand{\todocolorb}[1]{}
  \newcommand{\todocolorc}[1]{}
\fi
\newcommand{\mahmood}[1]{\todocolor{[[Mahmood: #1]]}}

\newcommand{\secref}[1]{Sec.~\ref{#1}\xspace}

\begin{document}

\title{Privacy-Preserving Collaborative Genomic Research: A Real-Life Deployment and Vision}

\author{Zahra Rahmani}
\authornote{Both authors contributed equally to this research.}
\email{zxr81@case.edu}
\author{Nahal Shahini}
\authornotemark[1]
\email{nxs814@case.edu}
\affiliation{%
  \institution{Case Western Reserve University}
  \city{Cleveland}
  \state{Ohio}
  \country{USA}
}

\author{Nadav Gat}
\authornote{Both authors contributed equally to this research.}
\email{nadavgat@mail.tau.ac.il}
\author{Zebin Yun}
\authornotemark[2]
\email{zebinyun@mail.tau.ac.il}
\affiliation{%
  \institution{Tel Aviv University}
  \city{Tel Aviv}
  \country{Israel}
}

\author{Yuzhou Jiang}
\email{yxj466@case.edu}
\affiliation{%
  \institution{Case Western Reserve University}
  \city{Cleveland}
  \state{Ohio}
  \country{USA}
}

\author{Ofir Farchy}
\email{ofir@lynx.md}
\affiliation{%
  \institution{Lynx.MD}
  \city{Palo Alto}
  \state{California}
  \country{USA}
}

\author{Yaniv Harel}
\email{yaniv10@tauex.tau.ac.il}
\affiliation{%
  \institution{Tel Aviv University}
  \city{Tel Aviv}
  \country{Israel}
 }

\author{Vipin Chaudhary}
\email{vxc204@case.edu}
\affiliation{%
  \institution{Case Western Reserve University}
  \city{Cleveland}
  \state{Ohio}
  \country{USA}
}

\author{Erman Ayday}
\authornote{Corresponding authors.}
\email{exa208@case.edu}
\affiliation{%
  \institution{Case Western Reserve University}
  \city{Cleveland}
  \state{Ohio}
  \country{USA}
}

\author{Mahmood Sharif}
\authornotemark[3]
\email{mahmoods@tauex.tau.ac.il}
\affiliation{%
  \institution{Tel Aviv University}
  \city{Tel Aviv}
  \country{Israel}
}

\renewcommand{\shortauthors}{Rahmani et al.}

\begin{abstract}
The data revolution holds a significant promise for the health sector. Vast amounts of data collected and measured from individuals will be transformed into knowledge, AI models, predictive systems, and digital best practices. One area of health that stands to benefit greatly from this advancement is the genomic domain. The advancement of AI, machine learning, and data science has opened new opportunities for genomic research, promising breakthroughs in personalized medicine. However, the increasing awareness of privacy and cyber security necessitates robust solutions to protect sensitive data in collaborative research. This paper presents a practical deployment of a privacy-preserving framework for genomic research, developed in collaboration with Lynx.MD, a platform designed for secure health data collaboration. 
The framework addresses critical cyber security and privacy challenges, enabling the privacy-preserving sharing and analysis of genomic data while mitigating risks associated with data breaches. By integrating advanced privacy-preserving algorithms, the solution ensures the protection of individual privacy without compromising data utility. A unique feature of the system is its ability to balance the trade-offs between data sharing and privacy, providing stakeholders with tools to quantify privacy risks and make informed decisions.
The implementation of the framework within Lynx.MD involves encoding genomic data into binary formats and applying noise through controlled perturbation techniques. This approach preserves essential statistical properties of the data, facilitating effective research and analysis. Additionally, the system incorporates real-time data monitoring and advanced visualization tools, enhancing user experience and decision-making capabilities.
The paper highlights the need for tailored privacy attacks and defenses specific to genomic data, given its unique characteristics compared to other data types. By addressing these challenges, the proposed solution aims to foster global collaboration in genomic research, ultimately contributing to significant advancements in personalized medicine and public health.
\end{abstract}


%\ccsdesc{Do Not Use This Code~Generate the Correct Terms for Your Paper}
%\ccsdesc[100]{Do Not Use This Code~Generate the Correct Terms for Your Paper}



%\received{2 July 2024}


\maketitle

\section{Introduction}
The world has an incremental progress in collecting genomic data over the years. The new generation of capabilities including AI, machine learning, and data science, provide a new opportunity to investigate and run research that may identify new variants, connect symptoms to root causes, and develop new (personalized) medicines. 
Every healthcare institution, a clinic, a health research center, and a medical organization, holds unique information that can highly contribute to the global research of specific cases and problems. 
While this is clear, at the same time, humanity increases the awareness and the perceived importance of privacy and cyber security. Making the maximum effort to protect the patients’ information and personal data of individuals that participate in clinical trials is of high importance. 
An important progress in defining the liability of these procedures’ organizers helps to protect the private data against unintentionally leakage or malicious attacks. The outcome of this progress between other reasons creates a situation that most of this valuable data stays at the organizations that gathered the information rather than data sharing and collaborative research across institutions. 

In this work, we provide our vision for privacy-preserving collaborative genomic research and showcase our first attempt for practical, a real-life deployment.  
In order to make this work practical and connected to real use-cases we work with an industry collaborator, Lynx.MD, and use the existing genomic data storage platform of this partner. Lynx.MD is an American Israeli start-up that has developed a platform for health data collaboration between healthcare institutions, pharma companies, and research foundations. The practical use-cases and the real data are essential for the ability to deeply study the research goals.
The proposed framework mainly addresses the cyber security and privacy issues that can better protect the data and reduce the damage in case of any data breach. It also allows researchers (users of Lynx.MD, who store their datasets in the platform) share their datasets among each other (e.g., for collaborative research) in a privacy-preserving way and share their AI models as black-box.
%The main motivation to gather the data into one space is to run AI models and smart algorithms to analyze the data. Part of the research will focus on exploring the risks and on reducing the potential attacks of the AI machines which is an evolving domain of attacks that is on the raise. Manipulating the AI models to get different results or to leak data out of the machine is a severe risk these days in such projects.

%ZAHRA: Modofied this paragraph:
%We also propose to have a component in the proposed system that shows the trade-off between sharing information and the privacy “cost” of doing this. These days there are no interpretable tools to measure the privacy loss, and as a result the decision about the data shared is not given in a conscious way. The individuals that are in charge of the data and should make the decisions are not always privacy and security experts, therefore the methodologies of presenting privacy penalty and providing the right background for making decisions are important part of the study.%
We also propose including a component in the system that demonstrates the trade-off between information sharing and the associated privacy costs. Currently, there are no interpretable tools available to measure privacy loss, resulting in decisions about data sharing being made without full awareness. Individuals responsible for these decisions are not always privacy and security experts, making it essential to develop methodologies that clearly present privacy penalties and provide the necessary context for informed decision-making.

Overall, this work addresses one of the most innovative domains that requires solving key challenges in order to make a dramatic change on the global ability to leverage distributed health data. Doing so will contribute to creating more collective data repositories that may enable the use of AI to find solutions, make diagnoses, and develop medicines for the most prioritized world diseases. To achieve our vision, we have already developed and integrated privacy-preserving algorithms with the Lynx.MD platform, created interpretable privacy risk tools, and deployed these mechanisms in high-performance computing environments. These steps ensure robust data protection and facilitate collaborative research. Moving forward, we will focus on evaluating and validating these tools with real-world data, fostering collaboration and training, and continuously enhancing cybersecurity measures. This roadmap aims to ensure robust data protection, facilitate collaborative research, and enable the practical use of AI and machine learning in genomic research while addressing privacy and security concerns. Note that although this work focuses on privacy, cybersecurity aspects are an integral part of protecting assets and data confidentiality. AI models are part of the protected landscape, along with the technologies used to better defend processes and data flow \cite{GANDAL2023103380,10.1145/3057729,10224927}. Future work will explore this aspect in greater depth.

\section{Background - Lynx.MD Platform}

Lynx.MD, founded by experts in healthcare, AI, and cyber security, unlocks the power of comprehensive healthcare data. The platform, powered by AI, large language models (LLMs), and computer vision, provides unparalleled access to real-world medical records, driving advancements in medical technology and the development of advanced diagnostics and therapies. 
%We ensure top-level security and data privacy, positively impacting millions through data-informed insights and groundbreaking research.

Lynx.MD surpasses traditional medical data platforms by incorporating a wide array of data beyond electronic medical records (EMR) and electronic health records (EHR). Its cloud-based technology centralizes and streamlines both structured and unstructured real-world patient data, including clinical notes, imaging, and video. This fosters data exploration and accelerates the development of data-driven healthcare applications while safeguarding patient privacy.

Key components of the platform include:
\begin{itemize}
    \item {\bfseries Real-World Data Access:} The platform provides expansive, multi-sourced data that is continuously updated and tokenized, ensuring researchers and medical professionals have the most current and relevant information available.

    \item {\bfseries Structured and Unstructured Data:} The platform offers access to a diverse range of data types, including images, videos, and physicians' notes. This comprehensive dataset reveals vital insights that are often missed by platforms that only handle structured data.

    \item {\bfseries Real-Time Live Data:} The platform ensures that data is always up-to-date, supporting continuous monitoring and allowing for customized patient data tracking. This real-time capability is crucial for timely and effective medical interventions.

    \item {\bfseries Unmatched Data Volume and Diversity:} The extensive repository of data supports thorough statistical analysis and generates precise insights. The diversity of data types and sources enhances the robustness of research outcomes and the development of innovative healthcare solutions.
\end{itemize}

%% The ``\verb|acmart|'' document class can be used to prepare articles
%% for any ACM publication --- conference or journal, and for any stage
%% of publication, from review to final ``camera-ready'' copy, to the
%% author's own version, with {\itshape very} few changes to the source.

\section{Related Works}
We discuss related work from four aspects: privacy risks for genomic data, privacy-preserving solutions for genomic research, machine learning (ML) and AI privacy, and law and policy.

\subsection{Privacy Risks for Genomic Data}

Genomic privacy has recently been explored by many researchers~\cite{erlich2014routes,survey:genomicera,ayday2013chills,amalio14,ayday_sp}. Several works have shown that anonymization does not effectively protect the privacy of genomic data~\cite{gitschier2009inferential,gymrek2013identifying,Hayden2013,MalinS04,anony:name,lin04,Kale2017AUM}. Previous research have demonstrated  that the identity of a participant of a genomic study can be revealed by using a second sample, that is, part of the DNA information from the individual and the results of the clinical study~\cite{related:homer,related:wang,im2012sharing,clayton2010inferring,Zhou_ESORICS_2011,related:shringarpureandbastumante,Rraisaro2016privacy,vonthenen2018reidentification}. Furthermore, several studies have examined phenotype prediction from genomic data, as a means of tracing identity~\cite{Humbert2015DeanonymizingGD,Lippert201711125,kayser2011improving,zubakov2010estimating,ou2012predicting,allen2010hundreds,manning2012genome,walsh2011irisplex,claes2014modeling,liu2012genome}. This line of research highlights the vulnerabilities in genomic data sharing, particularly when datasets are linked with other publicly available data sources. The predictive power of genomic data can be exploited to infer sensitive information about individuals, such as their susceptibility to certain diseases or traits like eye and hair color, which can then be used for discriminatory purposes.

Recent studies have also highlighted the potential for privacy breaches through familial search techniques, where an adversary can identify individuals by matching their genetic markers with those of their relatives~\cite{gymrek2013identifying}. This method leverages the genetic similarity among family members to breach privacy, making it a significant concern for genomic data sharing practices. The growing availability of public genetic databases intensifies this risk, as these repositories provide a rich source of genetic information that can be leveraged to cross-reference and identify individuals from anonymized datasets.

\subsection{Privacy-Preserving Solutions for Genomic Research}

Most proposed solutions to utilize or share genomic data are either based on obfuscation or the use of cyptographic techniques. Some researchers have proposed using differential privacy (DP) concept to mitigate membership inference attacks when releasing summary statistics, such as minor allele frequencies and chi-square values~\cite{differential:gwas,differential:gwas_yu,differential:gwas_johnson}. DP provides a mathematical framework that ensures the addition or removal of a single database item does not significantly affect the outcome of any analysis, thereby preserving the privacy of individuals in the dataset.

%Privacy guarantees of aforementioned DP-based solutions degrade due to existence of dependent records in a genomic database and this privacy risk elevates in dynamic databases. Existing works proposed general mechanisms to tackle this problem~\cite{liu2016dependence,song2017pufferfish,zhao2017dependent}.
%However, this privacy risk has not yet been studied for statistical genomic databases and existing mitigations fail to provide high data utility, which is a crucial requirement when releasing data from statistical genomic databases. 
%Similarly, during direct sharing of genomic data, LDP-based data sharing techniques do not consider the correlations in the shared data and they cause a significant loss in data utility.

A significant subset of cryptographic solutions focuses on private pattern-matching and comparison of genomic~\cite{Pastoriza_CCS_2007,DeCristofaro:2013:SGT:2517840.2517849,related:computation,related:securecomparison,Naveed:2014:CFE:2660267.2660291}. 
For privacy-preserving clinical genomics, Baldi \emph{et al.} proposed private set intersection-based techniques~\cite{related:baldi}. These techniques allow researchers to identify common genetic variants across datasets without revealing individual data points. Similarly, partially homomorphic encryption has been proposed for the privacy-preserving use of genomic data in clinical settings, enabling computations to be performed on encrypted data without revealing the underlying information~\cite{ayday_NDSS13,related:ermanclinic,ayday_Healthtech13,ayday_Globecom13}.
Kantarcioglu \emph{et al.} proposed homomorphic encryption-based techniques for privacy-preserving genomic research, which allow secure computations on genetic data without revealing the data itself~\cite{Kantarcioglu:2008:CAS:2222946.2223571}. 
%while Cassa \emph{et al.} proposed a cryptographic scheme to securely transmit externally generated sequence data\cite{related:transmission}
Wang \emph{et al.} proposed private edit distance protocols to find similar patients across multiple hospitals, enhancing collaborative research without compromising patient privacy~\cite{wang2015efficient}. These cryptographic methods offer robust privacy guarantees and are particularly suited for applications where sensitive genomic data must be processed securely. 

Despite these advancements, existing cryptographic solutions often face challenges related to interoperability and practicality. Different genomic data analysis problems require different cryptosystems, which can lead to significant security and efficiency issues when integrating multiple systems. For instance, the use of various encryption schemes for different purposes can create complexities in maintaining security and ensuring seamless data processing. Additionally, some cryptographic techniques can be computationally intensive, posing challenges for their practical implementation in large-scale genomic studies.

Future research must focus on developing unified frameworks that provide comprehensive privacy-preserving solutions while maintaining high data utility and computational efficiency. This includes designing cryptographic systems that are versatile and efficient enough to handle various genomic data analysis tasks. Moreover, it is essential to address the scalability of these solutions to ensure they can manage the increasing volume of genomic data generated by modern sequencing technologies. By advancing these areas, we can ensure that genomic research can progress without compromising the privacy and security of individual data.
%However, existing cryptographic solutions have certain drawbacks. Most importantly, they each consider and solve a different genomic data analysis problem using different cryptosystems. Thus, their interoperability and practicality are limited since using them together creates significant security and efficiency problems.

\subsection{ML/AI Privacy}\label{subsec:adversarial}

A wide range of privacy attacks have been proposed against ML models
in recent years, demonstrating various ways in which ML may leak
sensitive data during or after training~\cite{Carlini19Mem,
  Fredrikson15Inv, Shokri17MemInf, Zhu19GradLeak}. For example,
membership inference attacks enable adversaries with access to a
trained model to determine whether a record of data was used in
training~\cite{Shokri17MemInf}. As another example, model inversion
allows adversaries to reconstruct training samples from trained
models~\cite{Fredrikson15Inv}. Other attacks include, but are not
limited to unintended memorization~\cite{Carlini19Mem} and input and  
label reconstruction in federated learning~\cite{Zhu19GradLeak}.

ML privacy attacks can serve as a means to empirically quantify
leakage of private data under various settings. Still, to our
knowledge, these attacks have primarily been explored and used by ML
privacy academics. In our vision (\secref{sec:vision}),
the attack outcomes would be made available to the stakeholders on the
Lynx.MD platform to aid them in decision making when they need to decide
whether to release a model trained on the platform. We also note that
ML privacy attacks have been mainly studied in the vision and text
domains, with little effort exploring their effectiveness on genomic
and health data. Accordingly, it remains unclear whether
established attacks can reliably assess privacy leakage in ML in these
domains, suggesting that new attacks tailored for genomic and health
data may need to be developed.
%% \mahmood{Hence, it is important to assess whether these
%%   attacks work in the genomic and health domains, and if necessary,
%%   develop attacks that account for the domains' unique characteristics to
%%   enable reliable measurements. Indeed, our evaluation (Sec.~\ref{})
%%   demonstrate that leading membership-inference attacks previously
%%   found effective in the vision and text domains fail when tested
%%   against genomic models.}

Researchers have also proposed various methods to enhance ML
privacy~\cite{Abadi16DPSGD, Bonawitz17SecAgg, Nasr18MemDef, Papernot18Pate}. While some of these methods do not provably guarantee
privacy but have been demonstrated empirically effective~\cite{Nasr18MemDef},
other methods also carry provable guarantees~\cite{Abadi16DPSGD,
  Papernot18Pate}. Notably, 
the differentially private stochastic gradient descent
algorithm~\cite{Abadi16DPSGD} enables training ML models via
stochastic gradient descent while satisfying
differential privacy guarantees, thus, in a sense, providing plausible
deniability about whether certain records were used in training. In
our vision, these defenses would be provided as a service to users of
the Lynx.MD platform, thus helping protect data privacy when training
models (\secref{sec:vision}).

\mahmood{do we want to have a subsection on privacy dashboards?}

\subsection{Law and Policy}

Issues about the privacy and healthcare data are mostly covered by Health Insurance Portability and Accountability Act (HIPAA) and General Data Protection Regulation (GDPR) requirements~\cite{CFR2020}. 
The HIPAA Privacy Rule governs the disclosure of electronic and hard-copy medical information and allows patients to control their medical records to a degree. The Privacy Rule establishes that, with some exceptions, covered entities must obtain patients’ permission before disclosing their medical data to others \cite{CFR160103}. However, the HIPAA Privacy Rule only covers protected health information (PHI), which is “individually identifiable health information” that is electronically or otherwise transmitted or maintained \cite{HHS2012}. Thus, the rule does not cover de-identified information. 
%In our previous work, we showed that de-identification of (especially medical) information does not imply privacy, and we will explore this further in this research. 

HIPAA and the department of Health and Human Services provide detailed guidance as to how to de-identify records for purposes of the regulatory exemption. They can be de-identified by experts using reliable techniques such as suppression, generalization, and perturbation \cite{CFR2017}. In the alternative, the regulations list 18 identifiers that should be removed to achieve anonymization \cite{NCVHS2007}. Nevertheless, experts have determined that even with redaction of all of these identifiers, there is a small chance that skilled attackers could re-identify records. They could do so by matching de-identified data to publicly available information, such as voter registration records or news stories about individuals with illnesses or injuries \cite{Hoffman2012,CookeBailey2018}. Furthermore, multi-modal data that includes a wide variety of correlations, can more easily be re-identified than other anonymized data. 
%, as we have shown in our earlier research. 
It is also noteworthy that the HIPAA Privacy Rule contains many other exceptions that allow disclosure of PHI without patient authorization. Examples are disclosures for treatment, payment, and health care operations, as well as for law enforcement or public health purposes.

The General Data Protection Regulation (GDPR), on the other hand, provides more extensive protections and rights to data subjects. The GDPR protects individuals with regard to the “processing” of “personal data”, including personal data related to health. The GDPR applies to organizations outside of the EU, so long as they process data pertaining to EU citizens and the processing activities relate to the offering of goods or services or the monitoring of behavior that takes place within the EU. Thus, U.S. entities may have to comply not only with HIPAA, but also with the GDPR.
The GDPR protects personally identifiable data and pseudonymized data. The law does not cover “anonymous” data. Pseudonymised data are data that cannot be attributed to a specific data subject without additional information that is kept separate and is subject to technical and organizational safeguards to prevent re-identification of the data subject
%(cf. Article 4 (5) of the GDPR). 
“Anonymous” data are data that are assumed to be unlinkable to the data subject. Experts, however, may be skeptical that any data are truly anonymous in the healthcare sector.  In addition, like HIPAA, the GDPR carves out several exceptions that allow data use absent patient consent in the contexts of judicial proceedings, legal requirements, public health measures, and medical treatment.  

Thus, despite the existence of the HIPAA and GDPR, patients may continue to be worried about the privacy of their information when it is curated for research and AI/ML purposes. They may not understand the concept of de-identification or they may be aware of the risk of re-identification.  Moreover, they may simply wish to maintain control over their health data in the belief that it is their property and for the sake of autonomy.
%Therefore, we aim to develop solutions that will (i) give data donors full control of their data and (ii) provide them with privacy-risk quantification tools to show them the privacy implications of their data sharing decisions. Focusing on cross border data sharing, we will explore the compliance of our proposed enablers with HIPAA and the GDPR. Furthermore, as more and more data are curated for secondary uses, it is vital to develop new privacy and autonomy preserving tools. This is our main aim in this ETAI module. 

\section{Our Vision}
\label{sec:vision}

%Proposed privacy-preserving solution for integrating the solution into the Lynx platform >> ZAHRA

To address the critical challenges associated with genomic data privacy, we developed a comprehensive privacy-preserving solution for integration with the Lynx.MD platform. This solution leverages advanced privacy-preserving solutions to ensure robust privacy protections while maintaining the utility of shared genomic data.

In our broader vision, Lynx.MD serves as a sandbox platform where researchers can securely upload and share their genomic datasets. Our privacy preserving algorithms are seamlessly integrated into Lynx.MD, providing users with tools to analyze and share data without exposing sensitive information. Additionally, a user-friendly interface and sophisticated visualization tools will be developed to allow users to easily manage their datasets and understand the privacy and utility levels of shared data. These tools are crucial for making informed decisions about data sharing and analysis.

%>>ZAHRA
\subsection{Data Sharing Between Lynx.MD Users}

We chose to use a state-of-the-art privacy-preserving genomic dataset sharing algorithm \cite{jiang2023}. In this approach, genomic data is encoded into a binary matrix. Using a previously developed XOR mechanism, controlled noise is added to the data, effectively masking the original genomic sequences while retaining essential statistical properties. Following the noise addition, a novel post-processing technique is employed. This technique uses publicly known statistical data about the genomic datasets to enhance data quality without compromising the privacy ensured in the first stage. We implemented and deployed this privacy-preserving dataset sharing mechanism in a high-performance computing (HPC) environment to ensure it meets the scalability and memory requirements necessary for handling large genomic datasets.

In highlevel, researchers upload their genomic datasets to the Lynx.MD platform, which acts as a controlled environment for data sharing. We have conducted extensive testing to ensure that our privacy-preserving mechanisms are correctly applied to each dataset uploaded, safeguarding against potential privacy breaches. 
Once datasets are uploaded, the Lynx.MD platform implements the two-stage privacy-preserving scheme detailed earlier. This implementation is crucial for maintaining the confidentiality of the data while enabling its use for research purposes. The platform supports collaborative studies by allowing secure access to shared datasets and tools necessary for joint research efforts. 
%Researchers can access shared ML/AI models and research outputs, facilitating a collaborative and productive research environment.

By integrating this functionality on the Lynx.MD platform, we ensure that genomic data can be utilized for advancing personalized medicine and other research initiatives without compromising the privacy of the individuals represented in the datasets. This method aligns with our high-level goal of fostering collaboration while protecting sensitive information.

\subsection{Privacy Assurance in ML Models on the Lynx.MD Platform}\label{ml_related}
% \mahmood{an alternative section title?}

A fundamental premise of the Lynx.MD platform is that training machine learning (ML) models on the data available through the platform enables technologists and scientists to develop innovative technologies and derive valuable scientific insights. However, as previously discussed (\secref{subsec:adversarial}), releasing these models carries significant privacy risks. Adversaries may leverage access to the models to infer sensitive information about individuals' genomic or health data. Thus, it is imperative to ensure that the models leak minimal to no information about the training data before their release from the platform. To this end, our vision includes providing stakeholders (both data owners and users) with an automated evaluation system that assesses the extent to which models leak information about their training data. Additionally, stakeholders will receive recommendations for potential defenses that can be incorporated during training to enhance the protection of the training data.

Various privacy attacks have been proposed and thoroughly evaluated against ML models, as previously mentioned (\secref{subsec:adversarial}). However, genomic datasets exhibit significant differences compared to the image or text datasets commonly used in these evaluations~\cite{Shokri17MemInf, carlini_mem_inf}. First, genomic datasets typically have relatively small sample sizes, as collecting genomic data remains more expensive than collecting text or image data, which can be easily gathered from the internet, albeit not always labeled. Second, while there are millions of genomic features—such as Single Nucleotide Polymorphisms (SNPs)~\cite{Brookes1999}—that can be ingested by models, the number of useful features in a given dataset is often substantially smaller, with each feature admitting a limited set of values. In contrast, image and text datasets contain significantly more dimensions (e.g., 3,072–268,203 dimensions for standard image datasets), each admitting 256 different values~\cite{Shokri17MemInf}. These differences can impact the effectiveness of attacks. For instance, some attacks require auxiliary datasets to train surrogate models that are later used to infer membership~\cite{carlini_mem_inf}. These may be ineffective against genomic models due to the scarcity of auxiliary data. Therefore, it may be necessary to develop novel attacks specifically tailored for genomic data, and health data in general, to reliably assess ML models' leakage.

% Our findings (\todo{\secref{}}) evidence that 
% novel attacks for these domains may indeed be necessary, at least for
% membership inference. We intend to tackle these challenges in future work.

\subsection{Communicating the Privacy Risk to Users}

% \mahmood{Turn this into section 4.1, as the main part of 
%   the vision? Then, we can talk about statistics related 
%   to data- and ML-model-sharing (current 4.1 and 4.2). WDYT?}

Once the risk of private information leakage due to data sharing or
ML model release is assessed, it is necessary to communicate the risk to
users to facilitate their decision making regarding whether to share data
or models. To this end, among others, it is crucial to communicate to users the
(1) potential implications of certain risks (e.g., membership 
inference);
(2) theoretical guarantees of countermeasures applied (e.g., 
differential privacy mechanisms);
(3) empirical assessment of data leakage (e.g., for potential success 
of membership inference);
(4) assumptions behind the attacks for which the risk is estimated
(e.g., whether the adversary has auxiliary data);
(5) potential tradeoffs between utility and privacy that can be attained
with various countermeasures.
We intend to devise a privacy dashboard that conveys such information to
stakeholders in an usable manner. Primarily, for usability, it is essential
to optimize accessibility and run time. For the former, we plan to
rely on established literature offering ways to describe the sophisticated 
theoretical guarantees of certain defenses~\cite{Nanayakkara23ExplainDP} and conduct user 
studies to find adequate means to convey metrics estimated by ML privacy 
attacks that would be accessible by stakeholders with a wide-range of 
backgrounds and
expertise. For the latter, we intend to explore means to enable prompt 
assessment of necessary metrics (e.g., by proposing more efficient 
attacks).

\section{Proof-of-Concept Implementation}
%ZAHRA

We integrated the privacy-preserving genomic data sharing scheme into the Lynx.MD platform and rigorously assessed both the privacy and utility of the data using comprehensive metrics. Our analysis employed a variety of tests, including Average Point Error, Average Sample Error, and Mean and Variance Error, which together indicate how well the integrity and statistical properties of the original dataset are preserved post-transformation. The privacy level of the data was measured using the differential privacy parameter, $\epsilon$, which provides strong privacy protection at different levels. 

\begin{figure}[h]
  \centering
  \includegraphics[width=\linewidth]{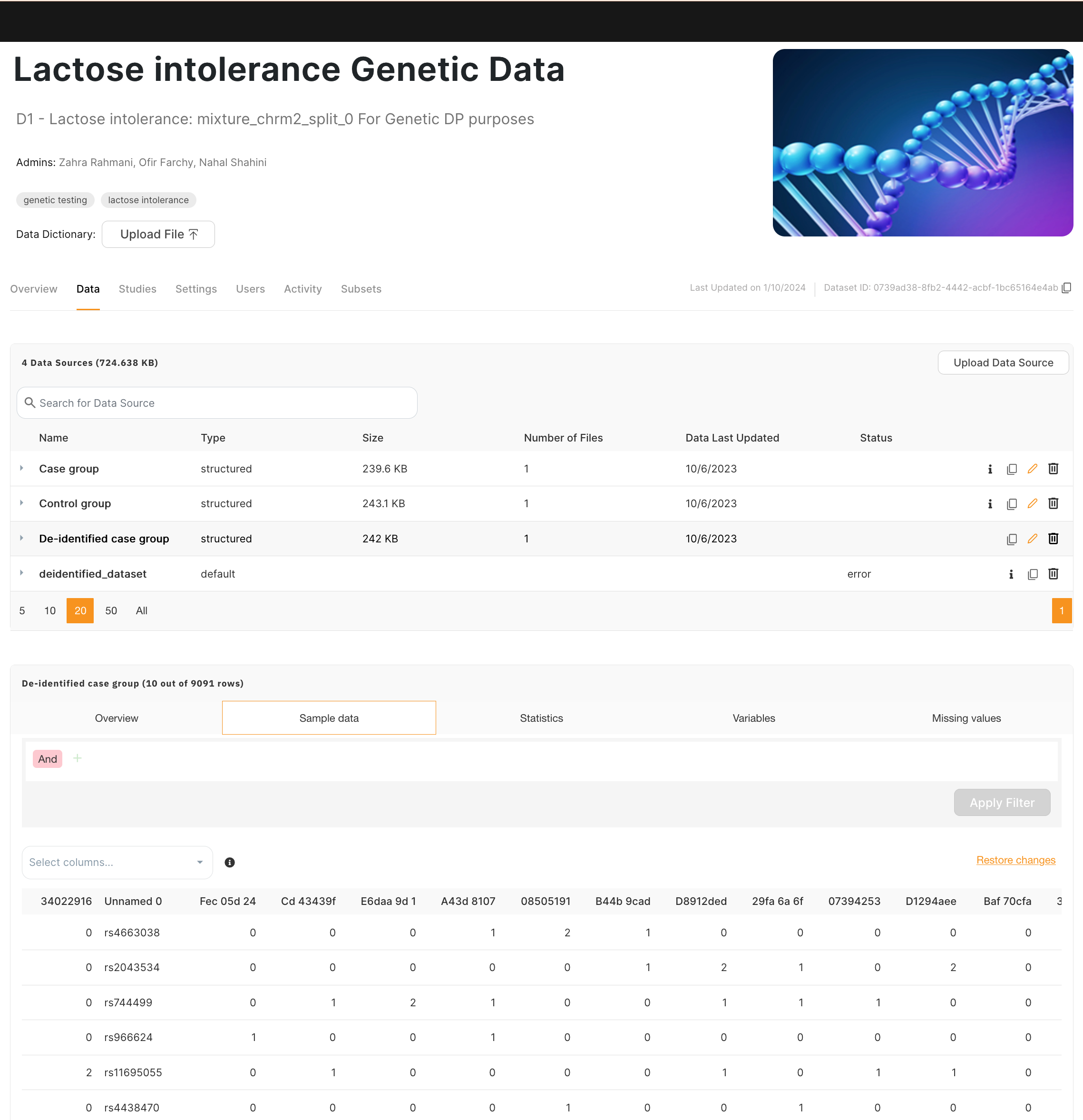}
  \caption{Lynx.MD Platform}
  \label{fig:dataset}
\end{figure}

\subsection{Dataset}
%ZAHRA
The dataset utilized for evaluating our privacy-preserving solution comprises a comprehensive collection of genomic data typical of what might be used in advanced medical research. This dataset features a broad spectrum of SNP variations, representing a diverse genetic background to ensure the generalizability of our results. Prior to the application of our privacy-preserving mechanisms, genomic data was encoded into a binary matrix format. This preparatory step was critical for facilitating the subsequent integration of our novel two-stage privacy-preserving algorithm, which applies controlled noise and utilizes publicly known statistics to enhance data quality.

We implemented and evaluated our proposed scheme on real-life genomic datasets from the OpenSNP project \cite{OpenSNP}, which is a public platform that allows users to share their genetic data, typically derived from consumer genetic testing services. We selected three phenotypes for our study: lactose intolerance, hair color, and eye color. The \textbf{lactose intolerance} dataset includes 9,091 SNPs from 60 individuals with lactose intolerance. The \textbf{hair color} dataset contains 9,686 SNPs from 60 individuals with dark hair. The \textbf{eye color} dataset is larger, with 28,396 SNPs among 401 individuals with brown eyes. Additionally, a \textbf{handedness} dataset is included with 28,396 SNPs among 401 individuals.

We built a reference dataset for each phenotype, aligning the SNPs with the target dataset to be shared. These reference datasets were constructed from the remaining data in the OpenSNP project. This thorough dataset selection and preparation ensured that our evaluation was robust and reflective of real-world scenarios. Note that we included these datasets on the Lynx.MD platform as users of the platform and ran the algorithms on the platform using these datasets as shown in Figure \ref{fig:dataset}.

\subsection{Utility and Privacy Evaluation}

Our results in \cite{jiang2023} demonstrate that the proposed scheme consistently outperforms existing methods in maintaining data utility, regardless of the privacy budget. This strong performance highlights how effective our two-stage privacy-enhancing mechanism is. By adding controlled noise and using a unique post-processing technique, we achieve an excellent balance between data privacy and usability. This is essential for moving forward with collaborative genomic research and personalized medicine.

On the other hand, the differential privacy framework safeguards against inference attacks, such as membership inference attacks, by injecting noise and protecting SNP value distributions, thereby offering robust privacy protection and high data utility. 
%Experimental evaluations affirm that this scheme outperforms existing methods like DPSyn and PrivBayes in terms of both accuracy and privacy, establishing it as a practical solution for real-world genomic dataset sharing

%We developed a novel approach to address the dual challenges of reproducibility and privacy in genomic research. We introduced a differential privacy-based scheme designed to enhance the sharing and validation of genomic datasets. Our method involved transforming these datasets into binary formats and applying XOR operations with binary noise that was calibrated based on the inherent correlations in the genomic data. To improve the utility of the perturbed data for genomic association validation, we integrated optimal transport theory to align the perturbed data more closely with the original distribution. Our evaluations on real genomic datasets demonstrated that our approach not only maintains individual privacy but also enhances data utility, offering substantial improvements over existing methods. This method proved particularly effective in detecting computational errors in genome-wide association studies (GWAS), thereby ensuring more robust and reliable research outcomes while minimizing privacy risks.

\subsection{System Performance}

The proposed privacy-preserving dataset sharing scheme introduces minimal overhead and demonstrates lower computational complexity compared to existing methods. Utilizing an efficient perturbation technique based on the XOR mechanism, the scheme significantly reduces time complexity by calibrating noise through column-wise correlation of SNPs, thus expediting the perturbation process without compromising privacy guarantees. This enhancement allows the scheme to handle large genomic datasets \cite{proserpio_14} practically within a reasonable timeframe. 

The scalability of the proposed dataset-sharing solution, when integrated with Lynx.MD, is highly promising. Designed to unlock the power of comprehensive healthcare data, Lynx.MD's robust infrastructure enables efficient management and distribution of large-scale genomic datasets. The integration leverages advanced data processing and security capabilities, ensuring that the solution can handle increasing volumes of data without performance degradation. The efficient perturbation and utility restoration mechanisms of the proposed scheme maintain low computational complexity, enhancing scalability as the dataset size grows. Consequently, the combined strengths of the platform and the scalable dataset-sharing solution facilitate effective and secure sharing of vast genomic data, driving significant advancements in healthcare research.

\section{Conclusion}
In this paper, we have demonstrated the feasibility and practicality of a privacy-preserving framework for collaborative genomic research, addressing the critical need for secure data sharing in the era of personalized medicine. By leveraging advanced privacy-preserving algorithms, our solution ensures robust protection of sensitive genomic data while maintaining its utility for research purposes.
The collaboration with Lynx.MD has been instrumental in validating our approach, showcasing how industry partnerships can enhance the deployment of privacy-preserving data platforms. Our method effectively balances the trade-offs between data sharing and privacy, providing stakeholders with transparent tools to assess privacy risks and make informed decisions.
Our experimental results confirm that the proposed framework outperforms existing methods in both privacy protection and data utility, highlighting its potential for broader application in genomic research and other fields requiring sensitive data handling. The integration of real-time monitoring and visualization tools further enhances the user experience, promoting more effective and secure collaboration.
Future work will focus on refining the privacy-preserving techniques and exploring additional applications in other domains. By continuing to address the unique challenges posed by genomic data, we aim to foster global collaboration and drive significant advancements in personalized medicine and public health.

\bibliographystyle{ACM-Reference-Format}
\bibliography{sample-base}


%%\begin{acks}

%%\end{acks}

%% \bibliographystyle{ACM-Reference-Format}
%% \bibliography{sample-base}

\appendix

\end{document}